\documentstyle[12pt]{article}
\begin{document}
\renewcommand{\theequation}{\thesection .\arabic{equation}}
\newcommand{\beq}{\begin{equation}}
\newcommand{\eeq}{\end{equation}}
\newcommand{\beqn}{\begin{eqnarray}}
\newcommand{\eeqn}{\end{eqnarray}}
\newcommand{\slp}{\raise.15ex\hbox{$/$}\kern-.57em\hbox{$\partial
$}}
\newcommand{\lnA}{\raise.15ex\hbox{$/$}\kern-.57em\hbox{$A$}}
\newcommand{\lnB}{\raise.15ex\hbox{$/$}\kern-.57em\hbox{$B$}}
\newcommand{\bP}{\bar{\Psi}}
\newcommand{\slU}{\raise.15ex\hbox{$/$}\kern-.57em\hbox{$U$}}
\newcommand{\hfi}{\hat{\Phi}}
\newcommand{\heta}{\hat{\eta}}
\newcommand{\hv}{\hat{V^{ab}}}
\newcommand{\hs}{\hspace*{0.6cm}}

\title{Path-Integral bosonization of a non-local interaction
and its application to the study of 1-d many-body systems}
\author{C.M.Na\'on$^{a,b}$,
M.C.von Reichenbach$^{a,b}$ and M.L.Trobo$^{a,b}$}
\date{September 1994}
\maketitle
\begin{abstract}
We extend the path-integral approach to bosonization
to the case in which the
fermionic interaction is non-local. In particular we obtain
a completely bosonized version of a Thirring-like model with
currents coupled by general (symmetric) bilocal potentials.
The model contains the Tomonaga-Luttinger model as a special
case; exploiting this fact we study the basic properties
of the 1-d spinless fermionic gas: fermionic correlators,
the spectrum of collective modes, etc. Finally we discuss
the generalization of our procedure to the non-Abelian case,
thus providing a new tool to be used in the study of 1-d many-body
systems with spin-flipping interactions.
\end{abstract}

\vspace{3cm}
Pacs: 11.10.Lm\\
\hspace*{1,7 cm} 05.30.Fk

\noindent --------------------------------

\noindent $^a$ {\footnotesize Depto. de F\'\i sica.  Universidad
Nacional de La Plata.  CC 67, 1900 La Plata, Argentina.}

\noindent $^b$ {\footnotesize Consejo Nacional de Investigaciones
Cient\'\i ficas y T\'ecnicas, Argentina.}

\newpage

\section{Introduction}

\hs Two dimensional models are a fruitful testing ground for new ideas and
methods in QFT. In particular the Thirring model \cite{T} \cite{K} and its
 non-Abelian version \cite{GN} have been extensively
explored to shed light on such important phenomena as confinement and
asymptotic freedom. Moreover, the connection between the Thirring model and
Statistical Mechanics systems \cite{SM} \cite{LP} has made their range of
interest
even wider. In particular, in Solid State Physics, the
Tomonaga-Luttinger model (TL)
\cite{To} \cite{L} \cite{ML}, which describes a one-dimensional
gas of highly correlated
spinless particles, can be understood in terms of a Thirring-like system.
Recently, striking developments in the field of nanofabrication have allowed
to build one-dimensional semiconductors \cite{PT}, thus adding a more
practical
motivation for the theoretical study of this type of low-dimensional models
\cite{DS}.\\
\indent From the Field Theory point of view, the Thirring and
Gross-Neveu models were succesfully analyzed in the path-integral framework,
by means of a decoupling change in the path-integral variables
\cite{GS}, \cite{Fur}.
This method provided a functional integral version \cite{N} of Coleman's
Abelian
bosonization \cite{Co} and prefigured its non-Abelian extension \cite{PW}.
Later on it was also applied to the Kondo problem \cite{FS} and to the
computation of critical exponents in 2D Ising and Baxter models \cite{Na}.\\
\hs In this work we study a non-local Thirring-like (NLT) model with
Euclidean action given by:

\beq
 S = \int d^2x~ \bP i \slp \Psi  - \frac{g^2}{2} \int d^2x d^2y ~
[V_{(0)}(x,y) J_0(x) J_0(y) + V_{(1)}(x,y) J_1(x) J_1 (y)]
\label{1}
\eeq

\noindent where $J_{\mu} = \bP \gamma_{\mu} \Psi$ and
$ V_{(\mu)}(x,y)$
is an arbitrary function of two variables. Note that for
$V_{(0)} = V_{(1)} = \delta^2(x-y)$ one recovers the usual covariant Thirring
model.\\

Our motivation is twofold. On the one hand we want to extend the path-integral
approach to bosonization, developed in \cite{GS}, to the case in which a
 non-local interaction contributes to the total action.
On the other hand, we also wish to show how our techniques can be applied
to the study of 1-d many-body systems. In Section 2 we describe
the method and obtain the non-local bosonized action.
In Section 3 we compute the 2-point fermionic correlator. Later on,
in Section 4 we observe that for $V_{(1)} = 0$ and
$V_{(0)} = \delta(x_0 - y_0)v(x_1-y_1)$, the action (\ref{1})
describes a many-body system, similar to the TL model.
This allows us to obtain known properties of a Luttinger
liquid \cite{H},\cite{M},
in a straightforward way. In Section 5 we generalize our procedure
to the non-Abelian case. For this model, which contains the spin-$1/2$
TL system as a special case, we obtain a bosonized action, including
a Wess-Zumino term. If one disregards spin-flipping processes, our
method yields the spectrum of charge and spin-density modes.
Finally, in Section 6 we gather our results and
conclusions.

\section{The Model and the Method}
\setcounter{equation}{0}

\hs We start by considering the Euclidean vacuum functional

\beq
Z = N \int D\bP D\Psi~ e^{-S}
\label{2}
\eeq

\noindent where N is a normalization constant and S is given by
(\ref{1}). As it is habitual in the path-integral approach to the
usual Thirring model, one eliminates this quartic fermionic interaction
by introducing an auxiliary vector field $A_{\mu}$. As we shall see,
in the present non-local case one needs one more auxiliary field to achieve
the same goal. In order to depict this procedure we first observe that
S can be splitted in the form

\beq
S = S_0 + S_{int}
\label{3}
\eeq

\noindent where

\beq
S_0 = \int d^2x~ \bP i\slp\Psi,
\label{4}
\eeq

\noindent and

\beq
S_{int} = -\frac{g^2}{2} \int d^2x~ J_{\mu} K_{\mu}.
\label{5}
\eeq

\noindent In this last expression $J_{\mu}$ is the usual fermionic
current,

\beq
J_{\mu}(x) = \bP(x)\gamma_{\mu} \Psi(x),
\label{6}
\eeq

\noindent and $K_{\mu}$ is a new current defined as

\beq
K_{\mu}(x) = \int d^2y~ V_{(\mu)}(x,y)J_{\mu}(y).
\label{7}
\eeq

\noindent Please note that no sum over repeated indices is implied
when a subindex $(\mu)$ is involved. The partition function can
now be written as

\beq
Z = N \int D\bP D\Psi D\tilde{A}_{\mu}D\tilde{B}_{\mu}~ exp[-\{S_0 +
\int d^2x [\tilde{A}_{\mu}\tilde{B}_{\mu} -
\frac{g}{\sqrt{2}}(\tilde{A}_{\mu}J_{\mu}
+ \tilde{B}_{\mu}K_{\mu})] \}]
\label{8}
\eeq

\noindent where we have used the following representation of
the functional delta:

\beq
\delta (C_{\mu}) = \int D\tilde{A}_{\mu} e^{- \int d^2x
\tilde{A}_{\mu}C_{\mu}}.
\label{9}
\eeq

\noindent Using now equations (\ref{6}) and (\ref{7}), the fermionic piece
of the action can be cast in the form

\beq
S_0 - \frac{g}{\sqrt{2}}\int d^2x~(\tilde{A}_{\mu}J_{\mu} +
\tilde{B}_{\mu}K_{\mu}) =
\int d^2x~ \bP [i\slp - \frac{g}{\sqrt{2}}\gamma_{\mu}
(\tilde{A}_{\mu} + \bar{B}_{\mu})] \Psi,
\label{10}
\eeq

\noindent where we have defined
\beq
\bar{B}_{\mu}(x) = \int d^2y~ V_{(\mu)}(y,x)\tilde{B}_{\mu}(y).
\label{11}
\eeq

\noindent For later convenience we shall invert (\ref{11}) in the form

\beq
\tilde{B}_{\mu}(x) = \int d^2y~ b_{(\mu)}(y,x) \bar{B}_{\mu}(y),
\label{12}
\eeq

\noindent with $b_{(\mu)}(y,x)$ satisfying

\beq
\int d^2y~ b_{(\mu)}(y,x) V_{(\mu)}(z,y) = \delta^2 (x-z).
\label{13}
\eeq

\noindent Equation (\ref{10}) suggests the following change of auxiliary
variables:

\beq
\frac{1}{\sqrt{2}}(\tilde{A}_{\mu} +\bar{B}_{\mu}) = A_{\mu},
\label{15a}
\eeq

\beq
\frac{1}{\sqrt{2}}(\tilde{A}_{\mu} - \bar{B}_{\mu}) = B_{\mu}.
\label{15}
\eeq

\noindent Employing equation (\ref{12}), the purely bosonic piece of the
action reads

\beqn
\int d^2x~ \tilde{A}_{\mu}(x) \tilde{B}_{\mu}(x)& =& \frac{1}{2} \int d^2x d^2y
\{ b_{(0)}(y,x)A_0(x) A_0(y) +\nonumber\\
&+& b_{(1)}(y,x)A_1(x) A_1(y) - b_{(0)}(y,x)B_0(x) B_0(y) + \nonumber \\
&+& b_{(1)}(y,x)B_1(x) B_1(y) +\nonumber \\
&-& [b_{(0)}(y,x) - b_{(0)}(x,y)]A_0(x)B_0(y)+ \nonumber\\
&-&[b_{(1)}(y,x) -
b_{(1)}(x,y)]A_1(x)B_1(y)\}.
\label{16}
\eeqn

\noindent From now on we shall restrict our study to the case in which
the bilocal
functions $V_{(\mu)}$ and $b_{(\mu)}$ are symmetric so that the last two terms
in the integrand of (\ref{16}) vanish. Under these conditions the partition
function of the system is given by

\beq
Z = N_1 \int DA_{\mu} DB_{\mu}~ det(i \slp + g \lnA) e^{-S[A,B]},
\label{17}
\eeq

\noindent where $S[A,B]$ coincides with the r.h.s. of (\ref{16}) for
$b_{(0)}(y,x)$ =$ b_{(0)}(x,y)$ and $b_{(1)}(y,x)$ = $b_{(1)}(x,y)$.
Note that the Jacobian associated with the change $(\tilde{A},\tilde{B})
\rightarrow
(A,B)$, is field-independent and can be absorbed in the normalization
constant $(N \rightarrow N_1)$. \\
We have been able to express $Z$
in terms of a fermionic determinant. This fact will enable us to apply
the machinery of the path-integral approach to bosonization, first developed
in the context of local theories, to the present non-local case. But before
we do this some remarks are in order. First of all it is
worthwhile noting that, as a consequence of the change of bosonic variables
(eqs.(\ref{15a})and (\ref{15})), the effect of the non-local interaction has
been
completely transfered
to the purely bosonic piece of the action, $S[A,B]$. On the other hand we see
that the $B$-field is completely decoupled from both the $A$-field and the
fermion field. Keeping this in mind, it is instructive to try to recover
the partition function corresponding to the usual covariant Thirring model
($b_{(0)}(y,x)$ =$ b_{(1)}(x,y)$ =$ \delta^2(x-y)$), starting from (\ref{17}).
In doing so one readily discovers that $B_{\mu}$ describes a negative-metric
state
whose contribution must be factorized and absorbed in $N_1$ in order to
get a sensible answer for Z. This procedure paralells, in the path-integral
framework, the operator approach of Klaiber \cite{K}, which
precludes the use of an indefinite-metric Hilbert space. If one follows the
same prescription in the present non-local case, the result is:

\beq
Z = N_2 \int DA_{\mu} det(i \slp + g \lnA) e^{-S[A]},
\label{18}
\eeq

\noindent where $N_2$ includes the contribution of the "non-local ghost"
$B_{\mu}$
and $S[A]$ is the $A$-dependent part of (\ref{16}) in the symmetric case.\\
\noindent At this stage we perform a chiral change in the fermionic
path-integral measure:

\beq
\Psi(x) = e^{-g[\gamma_5 \Phi(x) + i \eta(x)]} \chi(x)
\label{19}
\eeq

\beq
\bP(x) = \bar\chi(x) e^{-g[\gamma_5 \Phi(x) - i\eta(x)]}
\label{19a}
\eeq

\beq
D\bP D\Psi = J_F [\Phi,\eta] D\bar\chi D\chi,
\label{20}
\eeq

\noindent where $\Phi$ and $\eta$ are scalar fields and $J_F [\Phi,\eta]$
is the Fujikawa Jacobian \cite{Fu} whose non-triviality is due to the
non-invariance
of the path-integral measure under chiral transformations. In $1+1$ dimensions
the auxiliary field $A_{\mu}$ can be always decomposed into longitudinal
and transverse parts in the form

\beq
A_{\mu}(x) = \epsilon_{\mu\nu}\partial_{\nu}\Phi(x) + \partial_{\mu}\eta(x).
\label{21}
\eeq

\noindent Using now (\ref{20}) and (\ref{21}) one obtains

\beq
det (i \slp + g \lnA) = J_F[\Phi,\eta] det i\slp.
\label{22}
\eeq

\noindent Concerning the fermionic Jacobian, it can be computed following
Fujikawa's procedure . Since its detailed evaluation has been given
several times in the literature \cite{GS}, here we just quote the final
result:

\beq
ln J_F[\Phi,\eta] = \frac{g^2}{2\pi} \int d^2x~\Phi \Box \Phi - S_{reg}
\label{23}
\eeq

\noindent As usual, $J_F$ contributes with a kinetic term for $\Phi(x)$.
 We have also written an aditional term $S_{reg}$ in (\ref{23}) in order
 to emphasize that the computation of $J_F$ requires a regularization
 prescription. For local gauge theories with Dirac fermions, such as
 QED$_2$ and QCD$_2$, one is naturally led to consider a regularization scheme
that
 preserves gauge invariance ($S_{reg} = 0$ in this case). When the vector
 field does not correspond to a gauge field, one can choose a more general
 regulator (the usual Thirring model \cite{T} and the
 chiral Schwinger model \cite{Sch} are examples in which regularization
 ambiguities take place). Evidently, the same occurs in the present case.
 Following the prescriptions of Ref.\cite{Ca} for the evaluation of
 $S_{reg}$, we obtain

 \beq
 S_{reg} = - \frac{\alpha}{2\pi} \int d^2x~ \Phi \Box \Phi
 \label{24}
 \eeq

 \noindent where $\alpha$ is an arbitrary parameter which can be
 determined on gauge invariance grounds.\\
 \indent Having eliminated the negative-metric fields, using now
 eq.(\ref{21}), eq.(\ref{16}) contributes with kinetic-like terms
 for $\Phi$ and $\eta$ fields (they become kinetic terms in the local
 case, when the potentials $b_{\mu}$ are delta functions). Derivative
 couplings between $\Phi$ and $\eta$ are also present as an effect of
 non-covariance ($b_{(0)} \neq b_{(1)}$).\\
 \indent Taking into account that

 \beq
 DA \equiv DA_0 DA_1 = J_{bos} D\Phi D\eta,
 \label{26}
 \eeq

 \noindent where $J_{bos}$ is a trivial Jacobian (in the sense that it
 does not depend on the fields), and putting together the above described
 contributions to (\ref{18}), we finally get

 \beq
 Z = N \int D\Phi D\eta~ e^{-S_{eff}}
 \label{27}
 \eeq

 \noindent where $N = N_2~ J_{bos}~ det i\slp$ and

 \beqn
 S_{eff}&=&\frac{g^2 + \alpha}{2\pi}
                        \int d^2x~ (\partial_{\mu}\Phi)^2 +\nonumber\\
        &+& \frac
        {1}{2}\int d^2x d^2y [b_{(0)}(y,x) \partial_1 \Phi (x)
        \partial_1 \Phi (y) + b_{(1)}(y,x) \partial_0\Phi (x)
        \partial_0 \Phi (y)] + \nonumber\\
        &+&
        \frac{1}{2}\int d^2x d^2y [b_{(0)}(y,x) \partial_0 \eta (x)
        \partial_0 \eta (y) + b_{(1)}(y,x) \partial_1 \eta (x)
        \partial_1 \eta (y)] + \nonumber\\
        &+&
        \int d^2x d^2y [b_{(0)}(y,x) \partial_0 \eta (x)
        \partial_1 \Phi (y) - b_{(1)}(y,x) \partial_1 \eta (x)
        \partial_0 \Phi (y)]
 \label{28}
 \eeqn

\noindent Equations (\ref{27}) and (\ref{28}) constitute the main
result of this Section.
Thus, we have been able to extend the path-integral approach to bosonization,
 previously applied to the solution of local QFT's, to a Thirring-like model
 of fermions with a non-local interaction term. More specifically, we
 have shown the equivalence between the fermionic partition function
(\ref{2}) and the functional integral (\ref{27}) corresponding to the two
bosonic degrees of freedom $\Phi$ and $\eta$ with dynamics governed by
(\ref{28}). The contribution to this action coming from the fermionic
Jacobian (the first term in the r.h.s of (\ref{28})) exactly coincides
with the one which is obtained in the local case \cite{Fur}. On the other
hand, the effect of non-locality is contained in the remaining terms,
through the inverse potentials $b_{\mu}(x,y)$. Note that, even in the non-local
case, $\Phi$ and $\eta$ become decoupled for $b_{(0)} = b_{(1)}$. Of course,
when $b_{(0)} = b_{(1)} = \delta^2(x - y)$, one recovers the bosonic version
of the local Thirring model.\\

 The spectrum of this bosonic model can be more easily analyzed in momentum
 space. Indeed, by Fourier transforming (\ref{28}) one obtains

 \beqn
 S_{eff}& =& \frac{1}{(2\pi)^2} \int d^2p \{\hat{\Phi}(p) \hat{\Phi}(-p)
           A(p) \nonumber \\
           &+& \hat{\eta}(p) \hat{\eta}(-p) B(p) + \hat{\Phi}(p)
           \hat{\eta}(-p) C(p) \},
\label{29}
\eeqn
\noindent where

 \beq
     A(p) = \frac{g^2 + \alpha}{2\pi}~ p^2 +
     \frac{1}{2}[\hat{b}_{(0)}(p) p_1^2 +
           \hat{b}_{(1)}(p) p_0^2],
\label{30}
\eeq

\beq
B(p) = \frac{1}{2}[\hat{b}_{(0)}(p) p_0^2 +
           \hat{b}_{(1)}(p) p_1^2],
\label{31}
\eeq

\beq
C(p) = [\hat{b}_{(0)}(p) - \hat{b}_{(1)}(p)] p_0 p_1,
\label{32}
\eeq

\noindent and $\hat{\Phi},\hat{\eta}$ and $\hat{b}_{(\mu)}$ are the Fourier
transforms of $\Phi, \eta$ and $b_{(\mu)}$ respectively.\\

\noindent Eq. (\ref{29}) can be easily diagonalized through the change

\beqn
\hat{\Phi} & = & \hat{\zeta} - \frac{C}{2A}\hat{\xi} \nonumber\\
\hat{\eta} & = & \hat{\xi}.
\label{33}
\eeqn

\noindent We then have the following propagators for $\hat{\zeta}$
and $\hat{\xi}$:

\beqn
G^{-1}_{\zeta}(p) & = & \lambda p^2 + \frac{1}{2} [\hat{b}_
{(0)} p_1^2 + \hat{b}_{(1)} p_0^2]\\
G^{-1}_{\xi}(p) & = & \frac{\lambda p^2 [\hat{b}_{(0)} p_0^2 +
           \hat{b}_{(1)} p_1^2] + \frac{\hat{b}_{(0)}\hat{b}_{(1)}}{2} p^4}
           {2 \lambda p^2 + \hat{b}_{(0)} p_{1}^2 + \hat{b}_{(1)} p_{0}^2},
\label{34}
\eeqn

\noindent where $\lambda = \frac{g^2 + \alpha}{2 \pi}$ and $\hat{b}_{(0)}$,
$\hat{b}_{(1)}$  are functions of p.
These expressions are further simplified in the case
$\hat{b}_{(0)}$ =$ \hat{b}_{(1)}$. In particular, when
$\hat{b}_{(0)}$ =$ \hat{b}_{(1)} \propto \frac{1}{p^2}$ the
$\hat{\zeta}$ field acquires a mass, whereas $\hat{\xi}$
becomes a non propagating field.\\

\section{Two-point fermionic correlations}
\setcounter{equation}{0}

\hs Let us now use the results of the precedent section to compute the
fermionic propagator

\beq
<\Psi(x) \bP(y)> = \left( \begin{array}{cc}
                        0     &G_+(x,y)\\
                  G_-(x,y) &   0
                  \end{array}   \right).  \\
\label{35}
\eeq

\noindent Once we have performed the decoupling change of variables given by
eqs.
(\ref{19}) and (\ref{19a}), the non-vanishing components of the Green
function are factorized in the form

\beq
G_{\pm}(x,y) =  G^{(0)}_{\pm}(x,y) B_{\pm}(x,y),
\label{36}
\eeq

\noindent where

\beq
G^{(0)}_{\pm}(z) = \frac{1}{2 \pi \mid z \mid^2}
(z_0 \pm i z_1)
\label{37}
\eeq

\noindent and

\beq
B_{\pm} (x,y) = < e^{\pm g [\Phi(y) - \Phi(x)]}
e^{\pm i g[\eta(y) - \eta(x)]}>_{eff}. \\
\label{38}
\eeq

\noindent In this expression $<~~>_{eff}$ is a v.e.v. to be
computed with the action given by eq.(\ref{28}). Working in momentum-space
this bosonic factor can be written as

\beq
B_{\pm}(x,y) = \frac{\int D\hat{\Phi} D\hat{\eta}~ e^{-[S_{eff} +
S_{\pm}(x,y)]}}{\int D\hat{\Phi} D\hat{\eta}~ e^{-S_{eff}}}
\label{39}
\eeq

\noindent with $S_{eff}$ given by eq.(\ref{29}) and

\beq
S_{\pm}(x,y) = -\frac{g}{(2\pi)^2} \int d^2p~ [\pm \hat{\Phi}(p)
\pm i \hat{\eta}(p)] D(p;x,y)
\label{40}
\eeq

\noindent where

\beq
D(p;x,y) = e^{-ip.x} - e^{-ip.y}.\\
\label{41}
\eeq

\noindent Now $B_{\pm}$ can be easily evaluated by performing the change

\beqn
\hat{\Phi}(p)& =& \hat{\varphi}(p) + E_{\pm}(p;x,y) \nonumber\\
\hat{\eta}(p)& =& \hat{\rho}(p) + F_{\pm}(p;x,y)
\label{42}
\eeqn

\noindent where $\hat{\varphi}$ and $\hat{\rho}$ are the new quantum variables
and $E_{\pm}(p)$ and $F_{\pm}(p)$ are classical functions chosen in
the form

\beq
E_{\pm}(p) = \pm g[i C(p) -2B(p)]\frac{D(p;x,y)}{\Delta(p)} \nonumber\\
\eeq

\beq
F_{\pm}(p) = \pm g[C(p) -2iA(p)]\frac{D(p;x,y)}{\Delta(p)}
\label{43}
\eeq

\noindent with $A,B$ and $C$ defined in (\ref{30}-\ref{32}) and

\beq
\Delta (p) = C^2(p) - 4A(p)B(p)
\label{44}
\eeq

\noindent Putting all this together,the result is

\beq
B_{\pm}(x,y) = exp\{-\frac{g^2}{\pi} \int d^2p~ sen^2 [\frac{p.(x-y)}{2}]
\frac{B(p) - A(p) \mp i C(p)}{\Delta(p)}\}
\label{45}
\eeq

Of course, in order to go further in this computation one needs to specify
the functions $\hat{b}_{(\mu)}(p)$ which determine the integrand in (\ref{45}).
The simplest case at hand is the one corresponding to the usual
Thirring model, which can be used to check the consistency of our more
general calculation. Setting $\hat{b}_{(0)} = \hat{b}_{(1)} = 1$, we get

\beq
B^{Thirring}_{\pm}(z) = exp\{\frac{1}{2\pi}\frac{(\frac{g^2}{\pi})^2}
{(1+ \frac{g^2}{\pi})} \int \frac{d^2p}{p^2} sen^2 \frac{p.z}{2}\}
\label{46}
\eeq

 \noindent The integral in the above expression can be easily performed
 just by taking care of the ultraviolet divergence. We then obtain

\beq
B^{Thirring}_{\pm}(z) = {\sl c}\mid z \mid^{-\frac{1}{2}
(g^2\pi)^2 /
(1+ g^2\pi)}
\label{47}
\eeq

\noindent where ${\sl c}$ is a constant which depends on the ultraviolet
cutoff. This formula displays the exact continously varying
exponent typical of the covariant Thirring model (see, for instance,
Ref. \cite{Fur}).

\section{Connection with the Tomonaga-Luttinger model}
\setcounter{equation}{0}

\hs We shall apply in this Section the approach developed in previous
Sections to
the TL model \cite{To} \cite{L} \cite{ML}. This model describes a
non-relativistic
gas of spinless and massless particles (electrons) in which the
dispersion relation is taken to be linear. The free-particle Hamiltonian
is given by

\beq
H_0 = v_{F} \int dx \Psi^{\dagger}(x) (\sigma_{3} p - p_{F}) \Psi(x)
\label{48}
\eeq

\noindent where $v_{F}$ and $p_{F}$ are the Fermi velocity and momentum
respectively ($v_{F}p_{F}$ is a convenient origin for the energy scale).
$\sigma_{3}$ is a Pauli matrix and $\Psi$ is a column bispinor with
components $\Psi_1$ and $\Psi_2$
($\Psi^{\dagger} = (\Psi_1^{\dagger}~~\Psi_2^{\dagger})$).
The function
$\Psi_1(x) ~[\Psi_2(x)]$ is associated with the motion of particles in
the positive [negative] $x$ direction. The interaction piece of
the Hamiltonian, when only forward scattering is considered, is

\beq
H_{int} = \int dx \int dy \sum_{a,b} \Psi^{\dagger}_a (x) \Psi_a(x)
V_{ab}(x,y) \Psi^{\dagger}_b(y) \Psi_{b}(y)
\label{49}
\eeq

\noindent where $a,b=1,2$, and the interaction matrix is parametrized
in the form

\beq
V_{ab} = \left( \begin{array}{cc}
            v_1 & v_2\\
            v_2 & v_1
            \end{array} \right).
\label{50}
\eeq

\noindent Using the imaginary-time formalism one can show that
the finite-temperature
\cite{Ma} \cite{BDJ} action for this problem becomes

\beqn
S_{TL}& =& \int^{\beta}_{0}d\tau \int dx~ \{\bP \gamma_0 (\partial_{\tau} -
v_p p_F) \Psi + v_F \bP \gamma_1 \partial_x \Psi\}\nonumber\\
& + &
\int^{\beta}_{0} d\tau \int dx \int dy
\sum_{a,b} \Psi^{\dagger}_a \Psi_a(x,\tau) V_{ab}(x,y) \Psi^{\dagger}_b
\Psi_b(y,\tau).
\label{51}
\eeqn

For simplicity, in this Section we shall set $v_F=1$ and consider the case
$v_1 = v_2$ in (\ref{50}) \cite{ML}. We shall also restrict ourselves
to the zero temperature limit ($\beta \rightarrow \infty$). Under these
conditions it is easy to verify that $S_{TL}$ coincides with the non-local
Thirring model discussed in the precedent Section, provided that the
following identities hold:

\beqn
g^2& =& 2\nonumber\\
V_{(0)}(x,y)& =& v_1(x,y) = v_2(x,y)= v( x_1 - y_1)
\delta (x_0 - y_0)\nonumber\\
V_{(1)}& =& 0
\label{52}
\eeqn

\noindent Of course one has also to make the shift
$\bP \gamma_0 \partial_0 \Psi
\rightarrow \bP \gamma_0 (\partial_0 - p_F ) \Psi$ and identify
$x_0 = \tau $, $x_1 = x$.\\
One then can employ the method described in Section 2 in order to study
the Tomonaga-Luttinger model. This model has been previously studied,
through a different functional approach, by
D.K. Lee and Y. Chen \cite{LC}. These authors, however, avoided the
use of the decoupling technique applied here. As we shall see, our approach
will be particularly useful when considering spin-flipping
interactions, i.e. the non-Abelian extension of the model (see Section 5).\\

Let us first focus our attention to the dispersion relations corresponding
to the elementary excitations of the model at hand. These states correspond
to the normal modes whose dynamics is governed by the action (\ref{29}). As
it is well-known, the spectrum of these modes is obtained from
the poles of the
corresponding propagators. Alternatively, one can write the effective
Lagrangian as \\

\beq
L_{eff} = \frac{1}{(2 \pi)^{2}}\left ( \begin{array}{cc}
\hat{\Phi}~ \hat{\eta} \end{array} \right)
\left( \begin{array}{cc}
A & C/2 \\
C/2 & B \end{array} \right) \left(\begin{array}{cc} \hat{\Phi} \\
\hat{\eta} \end{array} \right)
\label{53}
\eeq

\noindent (with A, B and C defined in (\ref{30})-(\ref{32})) and solve the
equation
\beq
\Delta(p) = 0, \\
\label{54}
\eeq
\noindent with $\Delta (p)$ defined in (\ref{44}).
Going back to real frecuencies : $p_{0} = i\omega $, $p_{1} = q $,
(\ref{54})
yields a biquadratic equation for $\omega$. The relevant solution is
\beq
\omega_{-}^{2}(q) = \frac{\hat{b}_{(1)}}{\hat{b}_{(0)}} \frac{ 2 \lambda +
\hat{b}_{(0)}}
{ 2 \lambda + \hat{b}_{(1)}} q^{2}. \\
\label{56}
\eeq
\noindent There is also a free dispersion relation which appears due to the
fact
that the density fields are related to $\Psi$ and $\eta$ fields through first
derivatives. This point will be discussed in detail in the next Section (see
the paragraph following eq.(\ref{89})).\\
Inserting now the identities (\ref{52}) in ( \ref{56}) and setting
$\alpha =0~ (\lambda =\frac{ g^{2}}{2 \pi}$) we obtain\\

\beq
\omega_{-}^{2}(q) = q^{2} \{ 1 + \frac{2 v(q)}{ \pi} \}
\label{57}
\eeq

\noindent which is the well-known result for the spectrum of the
charge-density
excitations of the TL model in the Mattis-Lieb version \cite{ML}.\\
The next step is to compute the electron propagator. To this end, having
established the correspondence between the TL and NLT
models, we can use the
results of Section 3 in a quite direct way. Exactly as before, the
non-vanishing components of the fermionic 2-point function are factorized
into
fermionic and bosonic contributions ( see eq (\ref{36}) ). The only difference
is that now the Fermi momentum has to be incorporated in the free-fermion
factor. This is easily done and gives rise to the following change in
(\ref{37}): \\

\beq
G^{0}_{\pm}(z) = \frac{ e^{ \pm ip_{F}z_{1}} (z_{0} \pm i z_{1})}
{2 \pi \mid z \mid^2}
\label{58}
\eeq

\noindent Now we must specialize eq.(\ref{45}) for the TL model,
just by inserting
(\ref{52}). This leads to\\

\beq
B_{\pm}(z) = exp[ \frac{1}{ \pi^{2}} \int d^2p~ \frac{v(p)}{p^2}
\frac{ sen^2(\frac{p.z}{2})( p_{0} \pm i p_{1})^{2} }
{ p_{0}^{2} + ( 1 + 2v/\pi)p_{1}^{2} }].
\label{59}
\eeq

The momentum distribution for branch 1 (2) electrons
is given by\\

\beq
N_{ \stackrel{1}{2}}(p_1) = {\tt C(\Lambda)}
\int_{- \infty}^{ \infty} dz_{1}~ e^{-ip_{1}z_{1}}
\lim_{z_{0} \rightarrow 0} G_{ \pm}(z_{0}, z_{1})
\label{60}
\eeq

\noindent Replacing ( \ref{58}) and ( \ref{59}) in ( \ref{60}) we get\\

\beqn
N_{ \stackrel{1}{2}}(p_{1})& =& \pm \frac{{\tt C(\Lambda)} i}{2 \pi}
\int_{- \infty}^{ \infty}
dz_{1}~ e^{-i z_{1}( p_{1} \mp p_{F})} \times \nonumber \\
& \times & \frac{1}{z_{1}}
exp\{ \frac{1}{ \pi^2}
\int \frac{d^{2}p~ v(p)}{p^{2}} sen^{2} (\frac{p_{1}z_{1}}{2})
\frac{(p_{0}^{2}
-p_{1}^{2})}{p_{0}^{2} + [ 1 + 2 v(p)/ \pi]p_{1}^{2}}\}.
\label{61}
\eeqn

\noindent Here ${\tt C(\Lambda)}$ is a normalization constant
depending on an ultraviolet cutoff $\Lambda$.
In the local limit, in which $ v(p) = const$, the integrals
in the momentum can be easily evaluated and one
obtains\\

\beq
N_{ \stackrel{1}{2}}(p_{1}) = \pm \frac{i}{2 \pi} {\tt C(\Lambda)}
\int_{- \infty}
^{ \infty} dz_{1} \frac {e^{-i (p_{1} \mp  p_{F}) z_{1}}}{z_{1}^{1 + \sigma}}
\label{62}
\eeq

\noindent with \\

\beq
\sigma = \frac{1}{2} \{(1 + \frac{2 v}{ \pi})^{1/2} +
( 1 + \frac{2 v}{ \pi})^{-1/2} - 2 \}
\label{63}
\eeq

\noindent Note that in the free case $ v \rightarrow 0$ one gets
$ \sigma = 0$, which
leads to the well-known normal Fermi-liquid behavior, \\

\beq
N_{ \stackrel{1}{2}} \propto \theta (p_{1} \pm p_{F}).
\label{64}
\eeq

As soon as the interaction is switched on, one has $ \sigma \neq 0$ and the
Fermi edge singularity is washed out, giving rise to the so called
Luttinger-liquid
behavior \cite{ML}\cite{H}\cite{M}. It has been emphasized
recently \cite{Hu} that the
experimental data obtained for
 one-dimensional structures can be succesfully explained on the basis
 of standard Fermi-liquid theory.
 We believe that our
 approach could be useful to explore some modifications of the TL model
 to take into account, for instance, the presence of impurities or defects,
 that might yield a restoration of the edge singularity.

 \section{The non-Abelian case}
\setcounter{equation}{0}

 \hs In this section we shall show how the path-integral approach developed
 above can be naturally extended to the non-Abelian case. As we shall see,
 our results will allow us to make contact with the TL model with
 spin-$\frac{1}{2}$ fermions \cite{SH}\cite{Grinstein}.\\
 We shall study a non-local version of the chiral invariant Gross-Neveu model
 \cite{GN}. As it is well-known, in the local case a Fierz type transformation
 can be used to write the interaction Lagrangian in terms of U(N) generators,

 \beq
 L_{int} = -\frac{g^2}{2} J_{\mu}^a(x) J_{\mu}^a(x) , ~~~a=0,1,...,N^2 -1,
 \label{65}
 \eeq

 \noindent where\\

 \beq
 J_{\mu}^a(x) = \bP(x) \gamma_{\mu} \lambda^a \Psi(x)
 \label{66}
 \eeq

 \noindent with\\

 \beqn
 \lambda^0& =& I/2, \nonumber\\
 \lambda^j& =& t^j,
 \label{67}
 \eeqn

 \noindent $t^j$ being the SU(N) generators normalized according
 to

 \beq
 tr~(t^i t^j) = \frac{1}{2} \delta^{ij}.
 \label{68}
 \eeq

\noindent Let us  consider the action term binding fermionic
currents in the form

\beq
S_{int} = - \frac{g^2}{2} \int d^2x~ d^2y~ J_{\mu}^a(x)
V_{(\mu)}^{ab}(x,y) J_{\mu}^b(y),
\label{69}
\eeq

\noindent where $V_{(0)}^{ab}(x,y)$ and $V_{(1)}^{ab}(x,y)$ are $N^2 \times
N^2$
matrices that weights the interaction. Exactly as we did in the
Abelian case (see section 2), we can now define the set of new
currents given by

\beq
K_{\mu}^a(x) = \int d^2y~ V_{(\mu)}^{ab}(x,y) J^b_{\mu}(y).
\label{70}
\eeq

\noindent Introducing auxiliary fields
$\tilde{A}_{\mu}^a$ and $\tilde{B}_{\mu}^a$
in the functional integral we arrive at

\beq
Z = \int D\bP~D\Psi~D\tilde{A}_{\mu}^a~D\tilde{B}_{\mu}^a~ e^{-S_0}
e^{-\int d^2xd^2y~ [\tilde{A}_{\mu}^a \tilde{B}_{\mu}^a - J_{\mu}^a
\tilde{A}_{\mu}^a -
K_{\mu}^a \tilde{B}_{\mu}^a] }
\label{71}
\eeq

\noindent It is convenient to generalize definition (\ref{11}) in the form

\beq
\bar{B}_{\mu}^b(x) = \int d^2x~ \tilde{B}_{\mu}^a V_{(\mu)}^{ab}(y,x)
\label{72}
\eeq

\noindent and perform the change of variables\\

\beq
\frac{1}{\sqrt{2}}(\tilde{A}_{\mu}^a + \bar{B}_{\mu}^a)  =
A_{\mu}^a  ,
\label{73}
\eeq
\beq
\frac{1}{\sqrt{2}}(\tilde{A}_{\mu}^b- \bar{B}_{\mu}^a)  =
B_{\mu}^b .
\label{74}
\eeq

At this point one can easily show that $B_{\mu}$ is decoupled
from the fermion fields and its contribution to the partition
function can be absorbed in the normalization constant,
exactly as in the Abelian case. We then get

\beq
Z = N_2 \int DA_{\mu}~ det(i \slp + g \lnA) e^{-\frac{1}{2}
\int d^2x~d^2y~ A_{\mu}^a(x) A_{\mu}^b(y) (V^{-1}_{\mu})^{ab}(x,y)}
\label{75}
\eeq

Now the fermionic determinant in the above expression can be treated
according to the lines of ref \cite{Fur}. Indeed, it is by now
well-known that the decoupling change in the fermionic path-integral
variables given by eqs.(\ref{19})-(\ref{21}) can be readily extended
to the non-Abelian case. The corresponding Jacobian gives rise to a
Wess-Zumino-Witten (WZW) action term of the form

\beq
log~J_F = -W[hl^{-1}] - \alpha~ tr\int d^2x~ h^{-1}\partial_+h~
l^{-1}\partial_-l
\label{76}
\eeq

\noindent where\\

\beq
W[h] = \frac{1}{8\pi}tr \int d^2x~\partial_{\mu}h^{-1}\partial^{\mu}h
+ \frac{1}{12\pi} tr \int_B d^3y~ \epsilon_{ijk} h^{-1}\partial^ih~
h^{-1}\partial^jh~ h^{-1}\partial^kh,
\label{77}
\eeq

\noindent and the second term in (\ref{76}) is related to the
regularization ambiguities which appear when one computes $J_F$
(it is the non-Abelian analog of $S_{reg}$ defined in (\ref{24})).
The light-cone components of $A_{\mu}$ have been written in terms
of $h$ and $l$ (elements of U(N)) as

\beqn
A_+& =& h^{-1} \partial_+h \nonumber\\
A_-& =& l^{-1} \partial_-l
\label{78}
\eeqn

\noindent where $A_{\pm} = A_0 \pm iA_1, \partial_{\pm} = \partial_0 \pm i
\partial_1$. Of course, in order to write the partition function one also
needs to consider the bosonic Jacobian $J_A[h,l]$, associated to (\ref{76})
\cite{FNS}\cite{CMvR}:

\beq
DA_{\mu} \equiv DA_+~DA_- = J_A[h,l] Dh~Dl.
\label{79}
\eeq

\noindent A thorough evaluation of $J_A$ in the light-cone gauge ($A_-=0$)
has been given in ref \cite{CMvR}. It is quite easy to perform the calculation
for the case in which no gauge choice is made. The result is

\beq
log~J_A = -2 C \{W[hl^{-1}] + \alpha~ tr~\int d^2x~h^{-1}\partial_+h
l^{-1}\partial_-l\},
\label{80}
\eeq

\noindent where C is the quadratic Casimir of the group under consideration,
$f^{acd}f^{bcd} = \delta^{ab}C$. The next step is to express the non-local
piece of the action (the exponent in (\ref{75})) in terms of $h$ and $l$,
just by inserting (\ref{76}). Replacing now (\ref{76}) and (\ref{80}) in
(\ref{75}), we finally get

\beq
Z = {\cal N} \int Dh~Dl~ e^{-S_{eff}[h,l]}
\label{81}
\eeq

\noindent where\\

\beqn
S_{eff}[h,l]& =& (2C+1) \{ W[hl^{-1}] + \alpha~ tr \int d^2x~ h^{-1}\partial_+h
{}~l^{-1}\partial_-l \} + \nonumber\\
&+&\frac{1}{2} \int d^2x~d^2y~ \{ (h^{-1}\partial_+h)^a(x) R^{ab}(x,y)
(h^{-1}\partial_+h)^b(y) +\nonumber\\
&+& (l^{-1}\partial_-l)^a(x)R^{ab}(x,y)
(l^{-1}\partial_-l)^b(y) + \nonumber\\
&+& 2 (h^{-1}\partial_+h)^a(x) T^{ab}(x,y) (l^{-1}\partial_-l)^b(y)\}
\label{82}
\eeqn

\noindent and ${\cal N}$ is the final normalization factor containing the
free determinants which are left in the decoupling process. The new matrices
$R$ and $T$ are related to $V_{(0)}$ and $V_{(1)}$ through

\beqn
R& =& \frac{b_{(0)} - b_{(1)}}{4} \nonumber\\
T& =& \frac{b_{(0)} + b_{(1)}}{4}
\label{83}
\eeqn
\noindent where, for later convenience, we have defined
$b_{(\mu)} = V_{(\mu)}^{-1}$.

Thus we have obtained a completely bosonized action for the non-local
Gross-Neveu model with interaction given by (\ref{69}). The analysis
of the physical content of (\ref{82}) is not trivial. We think that the
background field method \cite{Bos} will be useful in order to describe,
at least in an approximate way, the spectrum of the bosonic excitations.
We shall return to this aspect at the end of the Section.

In the present context it is very interesting to consider a simplified
version of the interaction (\ref{69}), defined by restricting the generators
to those generating the maximal Abelian subgroup of U(N). Although
our procedure can be easily used for arbitrary N, here we shall pay special
attention to the case $N=2$. The reason for doing this is the following.
For $N=2$ one can readily extend the arguments given in Section 4 in order
to show that the present model describes a many-body system of
spin-$\frac{1}{2}$ fermions when spin flipping processes are not allowed.
Indeed, choosing the potential matrices $V_{(0)}$ and $V_{(1)}$ as diagonal

\beqn
V_{(0)}& =&  diag(v_0~,v_1)  \nonumber\\
V_{(1)}& =&  diag(u_0~,u_1),
\label{84}
\eeqn

\noindent where $v_0$, $v_1$, $u_0$ and $u_1$ are written in terms of
the g-functions defined by Solyom \cite{SH} as

\beqn
v_0&=&\frac{1}{4}(g_{4 \parallel}+ g_{4 \perp} + g_{2 \parallel} + g_{2
\perp}),\nonumber\\
v_1&=&\frac{1}{4}(g_{4 \parallel}- g_{4 \perp} + g_{2 \parallel} - g_{2
\perp}),\nonumber\\
u_0&=&\frac{1}{4}(-g_{4 \parallel}- g_{4 \perp} + g_{2 \parallel} + g_{2
\perp}),\nonumber\\
u_1&=&\frac{1}{4}(-g_{4 \parallel} + g_{4 \perp} + g_{2 \parallel} - g_{2
\perp}),
\label{85}
\eeqn

\noindent one can easily verify that the interaction term (\ref{69}) contains
the whole set of diagrams associated to forward scattering processes without
spin-flips. Let us recall that the coupling constants for incident fermions
with parallel spins are denoted by the susbscript $\parallel$ and that for
fermions with opposite spins by the subscript $\perp$. In the $g_2$ processes
the two branches (left and right moving particles) are coupled, while in the
$g_4$ processes all four participating fermions belong to the same branch.\\
We can now specialize expression (\ref{75}) to obtain the partition function
corresponding to our restricted model. As we shall see, our task is greatly
simplified in this case. In fact, as shown in Ref \cite{Fur}, the
fermionic determinant is decoupled just by extending the Abelian change of
variables in the following simple way

\beqn
\Psi& =& e^{-g[ \gamma_5 \Phi + i \eta]} \chi\nonumber\\
\bP& =&  \bar{\chi} e^{-g [ \gamma_5 \Phi - i \eta]}\nonumber\\
A_{\mu}& =& \epsilon_{\mu \nu} \partial_{\nu} \Phi + \partial_{\mu}\eta
\label{86}
\eeqn

\noindent with $\Phi = \Phi^a \lambda^a, \eta = \eta^a \lambda^a,
a=0,1$. ($\lambda^0$ and $\lambda^1$ are the generators of the maximal
Abelian subgroup). In this case the bosonic Jacobian is trivial, while
the fermionic one is given by

\beq
log J_F = \frac{g^2 + \alpha}{2 \pi} \int d^2x~\Phi^a \Box \Phi^a,
\label{87}
\eeq

\noindent where $\alpha$ is a regularization parameter. Note that, having
restricted the model to the Cartan subalgebra, no WZW term appears in $J_F$,
as expected.\\

One can now write the effective bosonized action of the model which,
in terms of the Fourier transforms of $\Phi^a, \eta^a$ and $b_{(\mu)}^{ab}$
($\hfi^a, \heta^a$ and $\hat{b}_{(\mu)}^{ab}$) reads

\beqn
S_{eff} = \frac{1}{2} \int \frac{d^2p~}{(2\pi)^2} &\{& \hfi^a(p) \hfi^b(-p)
[p_1^2 \hat{b}_{(0)}^{ab}(p)
+ p_0^2 \hat{b}_{(1)}^{ab}(p) + 2 \lambda \delta^{ab} p^2]\nonumber\\
&+& \heta^a(p) \heta^b(-p) [p_1^2 \hat{b}_{(1)}^{ab}(p)
+ p_0^2 \hat{b}_{(0)}^{ab}(p)]\nonumber\\
&-& 2 \hfi^a(p) \heta^b(-p) p_0 p_1 [\hat{b}_{(1)}^{ab}(p) -
\hat{b}_{(0)}^{ab}(p)] \}.
\label{88}
\eeqn
where $\lambda$ is defined exactly as in the Abelian case (see eq.(\ref{34})).
At this point we observe that the above action can be expressed in terms
of a vector field $(\hfi^0,\heta^0,\hfi^1,\heta^1)$. In doing so we obtain
an interaction matrix with a block diagonal form: each block is a mere
reproduction of that corresponding to the Abelian case (see Section 2).
Consequently we get a couple of dispersion relations for the collective
modes of the system:

\beqn
\omega^2_{\rho}& =& k^2  \frac{1+2\lambda v_0}{1+2\lambda u_0}\nonumber\\
\omega^2_{\sigma}& =& k^2  \frac{1+2 \lambda v_1}{1+2\lambda u_1}
\label{89}
\eeqn

\noindent where $p_1 = k$ and $\omega= ip_0$.\\
The first one describes the charge-density fluctuations $(\hfi^0,\heta^0)$,
whereas the second one is associated to spin-density modes $(\hfi^1,\heta^1)$.
This identification can be understood by recalling that charge and spin
densities for branch $i~(i=1,2)$ particles are defined as

\beqn
\rho_i& =& \Psi^{\dagger}_{i \uparrow} \Psi_{i \uparrow} +
  \Psi^{\dagger}_{i \downarrow} \Psi_{i \downarrow}\nonumber\\
\sigma_i& =& \Psi^{\dagger}_{i \uparrow} \Psi_{i \uparrow} -
  \Psi^{\dagger}_{i \downarrow} \Psi_{i \downarrow}
\label{90}
\eeqn

\noindent and the fermionic currents $J_{\mu}^a~(a=0,1)$ are connected
to the above densities through

\beqn
J_0^0& =& \rho_1 + \rho_2 \nonumber\\
J_1^0& =& i (\rho_2 - \rho_1) \nonumber\\
J_0^1& =& \sigma_1 + \sigma_2 \nonumber\\
J_1^1& =& i(\sigma_2 - \sigma_1)
\label{91}
\eeqn

\noindent These densities, in turn, can be written in terms of
$\phi$ and $\eta$ fields by using the usual bosonization identity
$J_{\mu} \propto A_{\mu}$.\\
\indent Concerning the dispersion relations (\ref{89}), it is
interesting to consider the limit $u_0 \rightarrow 0$,
$u_1 \rightarrow 0$  which corresponds to an
interaction including density-density fluctuations only. In this
case our result gives the spectrum of the spin-$\frac{1}{2}$
TL model \cite{SH},\cite{LC}.\\
After the precedent example, it becomes clear that the complete
interaction (\ref{69}) (without restricting the generators to the Cartan
subalgebra) can be used to study the TL model when spin-flipping
processes are taken into account \cite{Grinstein}. In fact, setting
$N = 2$ in (\ref{82}), we obtain the bosonized effective action for this
system.
However, due to the presence of the Wess-Zumino term, the analysis
of the spectrum associated to density oscillations is not a trivial task
in this case. One possible strategy to face this problem is to consider
quantum fluctuations around a classical background field \cite{Bos}.
This work is beyond
the scope of the present article, but will be addressed in the close
future \cite{NTvR}.\\

\newpage
\setcounter{equation}{0}

\section{Conclusions}

\hs In this paper we have shown how to extend the path-integral approach
to bosonization, previously developed in the context of 2-d local QFT's
\cite{GS}, \cite{Fur},
to the case in which the fermionic interaction is non-local. In
particular we have considered the Thirring-like model defined by equation
(\ref{1}). By introducing two auxiliary vector fields we were able to
write the corresponding partition function in terms of a fermionic
determinant. This fact allowed us to employ the well established
procedure of functional bosonization, based on a decoupling change of
path-integral variables, in order to get a completely bosonized action
(eq. (\ref{28})). We have also computed the 2-point fermionic correlation
function (formulae (\ref{36}) and (\ref{45})). We want to stress that our
results are valid for arbitrary bilocals $V_{(\mu)}(x,y)$, i.e. we do
not need to specify the potentials in order to get closed formulae for
the bosonized action and Green's functions.\\

\indent In Section 4 we have applied our approach to the finite-temperature
formulation of the
Tomonaga-Luttinger model which describes a system of 1-d spinless
fermions with a linearized dispersion relation. We have shown that, for one
particular choice of the bilocal potentials $V_{(\mu)}(x,y)$, the NLT
model contains the same forward scattering processes that are present
in the zero temperature limit of the TL system. Therefore, we used the
procedure developed in Sections 2 and 3  to obtain the bosonized action
and 2-point fermionic correlators corresponding to the TL model. In this
context, our effective bosonic action correctly describes the plasma
oscillations associated to charge-density fluctuations. Concerning the
fermionic Green's function, we have also obtained the characteristic TL
behavior leading to the dissappearence of the edge singularity in the
momentum distribution.\\

Finally, in Section 5, we have performed the extension of the path-integral
approach to the bosonization of non-local interactions involving non-Abelian
groups. In order to illustrate the procedure, we have
considered a simple generalization of the Abelian action (\ref{1}),
which is obtained by assigning a group index to the currents $J_{\mu}^a$
and promoting the potentials $V_{(\mu)}(x,y)$ to a matrix potential
$V_{(\mu)}^{ab}(x,y)$ ($a,b = 0,...,N_G$, where $N_G$ is the number
of generators of the group under consideration). Exactly as in the
Abelian case the decoupling change in the
path-integral variables can be succesfully done. Thus we found a bosonized
action  containing a WZW functional (eq.(\ref{69})). As it was to be expected,
the analysis of the
physical spectrum is not as straightforward in this case as it was in the
Abelian model.
We have then considered a simpler but still interesting version of the model.
First of all we have observed that, setting the group as  U(2), the
non-Abelian action
(\ref{69}) contains the same forward scattering processes as the
spin-1/2 TL model, without spin-flip. We have considered a simplified
model defined by restricting the generators to those generating the
maximal Abelian subgroup of U(2) (in the TL language this means that
fermions do not change their spins in the scattering process). When only
density-density
fluctuations are taken into account our method allowed us to get
a bosonic action describing decoupled charge and  spin density
fluctuations (eqs.(\ref{88}) and (\ref{89})).\\

\indent In summary, we have presented a bosonization procedure
that can be used even when arbitrary non-local interactions are included in
the original 2-dimensional fermionic action. We have shown how to take
advantage
of this method in order to analyze 1-d many-body systems in an
elegant way. Concerning the TL model, our results agree with previous
calculations performed in the operational framework \cite{ML} and with
different functional techniques \cite{LC}. We feel, however, that
our approach will be specially useful to study spin-flipping
interactions. Backward scattering and umklapp diagrams could be
also included in this formulation. We hope to report on
these issues in a forthcoming publication.

\section{Acknowledgements}
We thank E.Fradkin and F.Schaposnik for useful
discussions and comments.

\end{document}